# Properties of the low-frequency phonon spectra of ferroelectric barium titanate-based heterostructures

E. D. Gor'kovaya[a], Yu. A. Tikhonov[a], V. I. Torgashev[a], A. S. Mikheykin[a], I. A. Lukyanchuk[b], D. Mezzane[c], N. Ortega[d], A. Kumar[d], R. S. Katiyar[d], and A. G. Razumnaya[a,*]

[a]Faculty of Physics, Southern Federal University, Rostov-on-Don, Russia; [b]Laboratoire de Physique de la Matière Condensée, Université de Picardie Jules Verne, Amiens, France; [c]LMCN, Cadi Ayyad University, Marrakesh, Morocco; [d]Department of Physics and Institute for Functional Nanomaterials, University of Puerto Rico, San Juan, Puerto Rico, USA

## ABSTRACT

Polarized Raman spectra of the epitaxial $Ba_{0.5}Sr_{0.5}TiO_3$ film, bi-color $BaTiO_3/Ba_{0.5}Sr_{0.5}TiO_3$ superlattice and tri-color $BaTiO_3/Ba_{0.5}Sr_{0.5}TiO_3/SrTiO_3$ superlattice were studied in a broad temperature range of 80-700 K. Based on the temperature dependence of the polar modes we determined the phase transitions temperatures in the studied heterostructures. In the sub-THz frequency range of the $Y(XZ)\bar{Y}$ spectra, we revealed the coexistence of the Debye-type central peak and soft mode in bi-color $BaTiO_3/Ba_{0.5}Sr_{0.5}TiO_3$ superlattice.

## 1. Introduction

Ferroelectric $Ba_xSr_{1-x}TiO_3$ (BSTx) materials attract great interest for a wide range of applications due to such outstanding properties as tunable dielectric permittivity, high electro-optic coefficient, and high refractive index. They have important application potential in design of electro-optic modulators and switchers, for fiber-optic communication lines in the gigahertz frequency range and for tunable microwave devices, such as phase shifters, waveguides, steerable antennas, and ferroelectric varactors [1].

The $Ba_{0.5}Sr_{0.5}TiO_3$ (BST0.5) solid solution material considered for such applications undergoes the ferroelectric-to-paraelectric phase transition below room temperature [2]. It was established that the BST0.5 bulk ceramics exhibit the same phase transitions sequence as it was observed in bulk $BaTiO_3$ (BT): from cubic $m3m$ to tetragonal $4mm$ phase at ~230 K, then to orthorhombic $mm2$ phase at ~170 K, and finally to rhombohedral $3m$ phase at ~130 K [2, 3]. The phase transition sequences in thin films and superlattices based on the BSTx are usually different from that in pure bulk materials with the same compositions. The «temperature – misfit strain» theoretical phase diagram of the BT and BSTx films contains at least four phases proposed by Pertsev et al [4]: paraelectric $p$-phase at high temperatures, orthorhombic $aa$-phase at positive misfit strains, tetragonal $c$-phase at negative misfit strains, and monoclinic $r$-phase at low temperatures. Thus, for BSTx films epitaxially grown on (001)-oriented cubic substrate there are two paths of phase transitions as a function of a misfit strain: (i) $p - c - r$ at negative misfit strain, and (ii) $p - aa - r$ at positive misfit strain [5].

Ferroelectric thin films and artificial superlattices composed of alternating ferroelectric layers usually exhibit better functional parameters than bulk materials of the same compositions or even can have new properties, not reachable in the bulk. Recently, heterostructures consisting of alternating ferroelectric, paraelectric and ferromagnetic layers with different compositions have been widely investigated [6-12].

According to previous Raman and far-infrared studies, the BT single crystals exhibit a complicated combination of displacive and order-disorder behavior at ferroelectric transition [13]. Below phase transition to the ferroelectric phase, the polar modes arise in the Raman spectra due to displacements of polar ions, which cause a change in the crystal structure during the phase transition. As shown by Scalabrin *et al* [14] and Burns and Dacol [15], the overdamped ferroelectric soft mode was observed in the tetragonal phase of BT crystal with the frequency of 35±5 cm$^{-1}$ and half-width of 100±10 cm$^{-1}$ at room temperature. Besides the higher frequency polar soft mode, the lower frequency Debye-like central mode was observed in BT single crystals [16-18], in BST ceramics and in polycrystalline films [19, 20] in the sub-THz frequency range.

In this work, we are interested in the coexistence of the coupled soft mode provided by the intra-well vibrations of the Ti ions and the central mode relaxations corresponded to the Ti ions jumps between eight equal positions, described by the eight-site model [21]. We perform the comparative Raman studies of the epitaxial BST0.5 film, bi-color BT/BST0.5 and tri-color BT/BST0.5/ST superlattices deposited on the cubic (001)MgO substrates in a broad temperature range of 80-500 K. We discuss the temperature dependence of the polar phonons with the aim to understand the lattice dynamics and particularly the emergence of the Debye-like central mode and soft mode at the ferroelectric phase transitions.

## 2. Experimental

Epitaxial $Ba_{0.5}Sr_{0.5}TiO_3$ (BST0.5) thin film, bi-color artificial $BaTiO_3/Ba_{0.5}Sr_{0.5}TiO_3$ (BT/BST) and tri-color $BaTiO_3/Ba_{0.5}Sr_{0.5}TiO_3/SrTiO_3$ (BT/BST/ST) superlattices were grown on cubic single crystal (001)MgO substrates by pulsed laser deposition using alternating focusing of the laser beam on stoichiometric BT, BST, and ST targets. The bi- and tri-color superlattices were symmetric with the modulation periods of $\Lambda$ = 135 Å and 143 Å, consequently. The total thickness of the grown BT/BST and BT/BST/ST superlattices was 1000 nm, while that of the BST-film was 600 nm. For comparison, the $Ba_{0.5}Sr_{0.5}TiO_3$ (BST05) ceramics sample was prepared by conventional solid-state reaction method from stoichiometric mixtures of $BaCO_3$, $SrCO_3$, and $TiO_2$ raw materials.

The structural perfection of the thin films and superlattices, the unit cell parameters of the layers, the modulation periods $\Lambda$ in the superlattices were determined by X-ray diffraction using a Rigaku Ultima IV diffractometer (Cu$K_{\alpha 1}$ radiation, 2θ/ω).

The micro-Raman spectra were excited by the polarized radiation of an argon laser (λ = 514.5 nm) and were recorded by a Jobin Yvon T64000 spectrometer equipped with a CCD detector. The exciting radiation was focused on a sample, using an Olympus optic microscope; the focused laser beam diameter was 2 μm on the sample surface. The temperature-dependent Raman measurements were performed using TS1500 and FDCS 196 heating and freezing stages, respectively, providing temperature stability of ± 0.1°C. The polarized Raman spectra of BST0.5 film, BT/BST and BT/BST/ST superlattices were measured on the samples oriented exactly according to the crystallographic axes of the cubic substrate so that $X \parallel [100]$, $Y \parallel [010]$, and $Z \parallel [001]$. The registration of the polarized spectra was performed in the side-view backscattering geometry, where the incident light is parallel to substrate and polarization of incident and scattered light is perpendicular. Raman spectra were obtained in frequency range 10-1000 cm$^{-1}$. All recorded spectra were corrected by the Bose-Einstein temperature factor. Note, that the MgO substrate has no Raman lines in the studied spectral range.

## 3. Results and discussion

At high temperature, the BT single crystal has a cubic structure with space group *Pm3m* symmetry. In this case, the selection rules for $O_h$ point group predict the presence of triply degenerated three $F_{1u}$ and one $F_{2u}$-type vibrations. Thus, in the cubic phase, we may expect the triply degenerate irreducible represantations $\Gamma_{cub}=3F_{1u}+F_{2u}$. The $F_{1u}$ modes are IR active, whereas $F_{2u}$-type mode is neither IR nor Raman active. On cooling, the bulk BT undergoes a transition to tetragonal phase with space group *P4mm* ($C_{4v}$ point group). In this phase, each of $F_{1u}$-type vibrations split into Raman and IR active $A_1$ and E modes, while $F_{2u}$ vibration splits into $B_1$ and E modes. Thus, in the tetragonal phase, we expect $\Gamma_{tet}=4E+3A_1+B_1$ phonon modes. The $B_1$ mode is the only Raman active. Since polar $A_1$ and E modes are related to the molecule polarizability changes, the long-range dipole-dipole coupling splits polar $A_1$ and E modes into the longitudinal and transverse optical (LO and TO) components. Detailed correlation of the polar $A_1$ and E-type modes in the Raman spectra of the tetragonal BT crystal was performed in [14]. For epitaxial BST0.5 film, we obtained polarized Raman spectra in normal and side-view backscattering geometries to record pure $A_1$ and E modes using the micro-Raman setup.

The Raman spectra of the bulk BST0.5 ceramics were reported by Yuzyuk [22]. It was shown that at room temperature the BST0.5 ceramic sample is paraelectric and any vibration modes are forbidden in Raman spectra. Nevertheless, the room-temperature Raman spectrum of the BST0.5 ceramics contains two broad bands at frequencies of about 230 and 580 cm$^{-1}$ caused by the dynamic disorder of Ti atoms as in the case of pure BT crystal. On cooling, the Raman spectra for the bulk BST0.5 ceramics demonstrate the main characteristic changes during transitions into tetragonal, orthorhombic and rhombohedral phases. The sequence of the transitions is similar to that observed in the BT single crystal, but all transitions are shifted to lower temperatures. Moreover, the phase transitions in the BST0.5 ceramics are diffuse and the coexistence of phases is observed in some temperature ranges.

Next, we consider the temperature evolution of the crossed, $Y(XZ)\overline{Y}$, and diagonal, $Y(XX)\overline{Y}$, Raman spectra of BST0.5 film in a temperature range of 80-320 K (Fig. 1). It worth noting that at the room temperature the bulk BST*x* samples are tetragonal ($a = b \neq c$) for Ba concentration $x > 0.7$ [2] and cubic for $x < 0.7$. In the strained BST$x$ films, the degree of the tetragonal distortions can rise because of the two-dimensional stresses imposed by the substrate. Such epitaxial stresses induce the changes of the transition temperatures and the sequence of phase transitions. It is known that the BST0.8 film deposited on (001)MgO substrate undergoes ferroelectric phase transitions to the tetragonal phase with the polarization perpendicularly to the substrate (*c* domain) at 540 K [23], whereas in the BST0.8 solid solutions this transition is observed at about 340 K [2]. Unexpectedly high Curie temperature (~500 K) was revealed in BST0.4 film due to the effect of the interface-induced build-in electric fields [24]. According to our x-ray data the BST0.5 film is cubic at room temperature with lattice parameter $a = 0.3962$ nm, which is coherent with the Raman measurements. At room temperature, for paraelectric BST0.5 film, all optical modes are forbidden in the spectra and only two disorder-activated broad bands centered at 230 and 580 cm$^{-1}$ are observed.

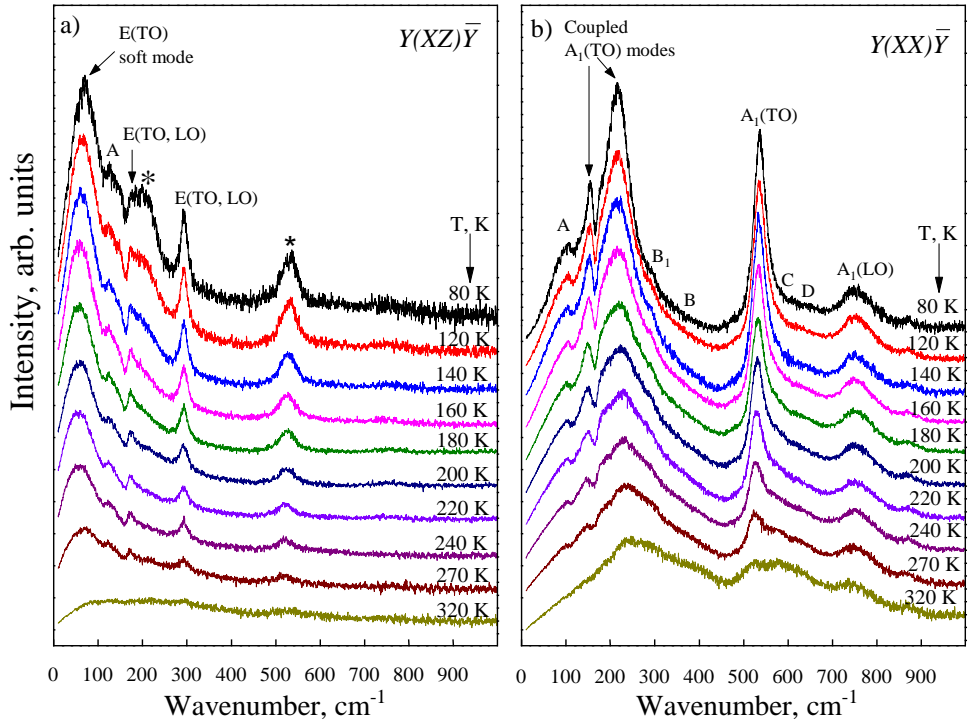

**Figure 1**. Raman spectra of the BST0.5 film in the (a) crossed $Y(XZ)\overline{Y}$ and (b) parallel $Y(XX)\overline{Y}$ scattering geometries at selected temperatures. The A$_1$-type phonon modes in $Y(XZ)\overline{Y}$ spectra, leaked from diagonal geometries due to depolarization are marked by the asterisks. The A, B, C and D bands are the disorder-induced lines.

On cooling, the polar modes arise in the $Y(XZ)\overline{Y}$ and $Y(XX)\overline{Y}$ spectra, their intensities increasing when the temperature decreases. At low temperatures, the polarized $Y(XZ)\overline{Y}$ and $Y(XX)\overline{Y}$ Raman spectra of the BST0.5 film contain all features typical for the BT single crystal [14].

With the aim to get the quantitative information about soft mode behavior, the experimental spectra were fitted using a sum of damped harmonic oscillators. The result of deconvolution of the $Y(XZ)\overline{Y}$ spectrum at 200 K is shown in Fig. 2a. Thus, we observe the E-type eigenmodes, characteristic for the crossed $Y(XZ)\overline{Y}$ spectra at frequencies 172, 291 and 747 cm$^{-1}$, and the underdamped soft mode E with the frequency 55 cm$^{-1}$ and the half-width of about 90 cm$^{-1}$. Compared with the data for the BT crystal, the significant shift of the soft mode frequency in the BST0.5 film is caused by two-dimensional stresses, induced by the substrate. Besides E modes, the $Y(XZ)\overline{Y}$ spectrum contains the leaked A$_1$-type bands from diagonal geometries at 213 cm$^{-1}$ and at 520 cm$^{-1}$. The last one overlaps the two very week E modes located at a frequencies of about 469 and 495 cm$^{-1}$. Moreover, the four disorder-induced A, B, C broad bands are observed in the $Y(XZ)\overline{Y}$ and $Y(XX)\overline{Y}$ spectra at frequencies 125, 360, 570, 600 cm$^{-1}$ due to intrinsic disorder of Ti ions.

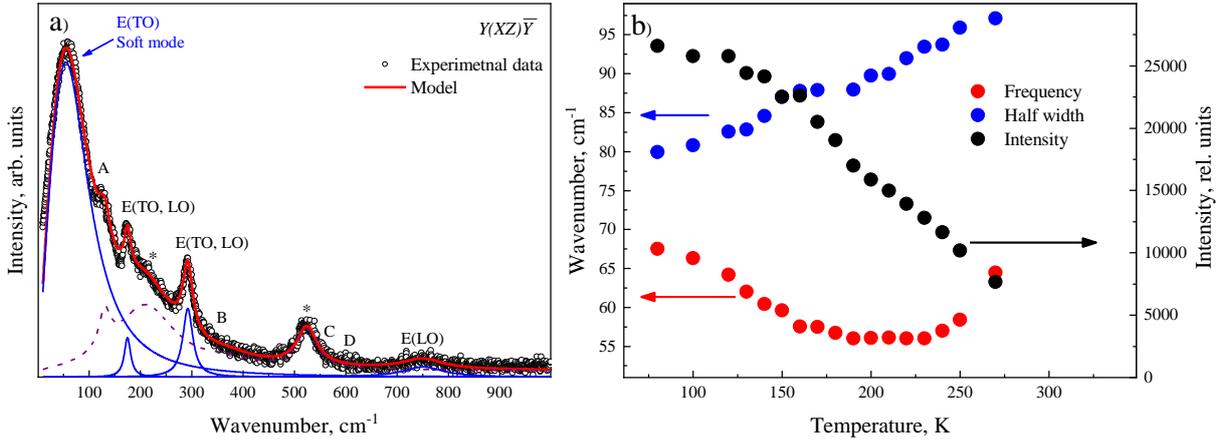

**Figure 2**. (a) Raman spectrum of the BST0.5 film in the $Y(XZ)\overline{Y}$ scattering geometry at 200 K and the result of deconvolution. The E-modes are denoted by the blue solid line. The A$_1$ phonon modes leak from diagonal $Y(XX)\overline{Y}$ geometries because of the depolarization and the disorder-activated A, B, C and D lines are marked by dashed lines. (b) Temperature dependences of integral intensity, frequency and half-width of the soft mode in the BST0.5 film.

As one can see in Fig. 1, there is some depolarization in the $Y(XZ)\overline{Y}$ spectra at 80 K. The leak of the A$_1$ bands into the crossed $Y(XZ)\overline{Y}$ scattering geometries may occur due to the presence of domain walls and dislocations and lowering of crystal symmetry. Note that the summarized contribution of the leaked A$_1$ lines in the $Y(XZ)\overline{Y}$ spectra becomes more significant as temperature decreases at about 200 K. Furthermore, the temperature evolution of the soft mode frequency changes in the slope and we observe a flat minimum in the wide temperature interval of 180-230 K (Fig. 2b). We assume that such transformations in the spectra may indicate a phase transition in the BST0.5 film. At room temperature, the misfit strain, determined from lattice parameters of the BST0.5 film is negative, $u_m = -0.7\%$. In this case, according to phenomenological theory for BST0.5 thin film [5], we expect the transitions from the paraelectric to tetragonal *c*-domain phase, and then, to the monoclinic *r* phase. Thus, above 270 K, we have found that Raman spectra of studied BST0.5 film correspond to the paraelectric phase. The polar mode parameters at higher temperature is impossible to determine since the corresponding peaks are hidden by the two broad disorder-induced lines. At about 270 K, the transition to ferroelectric tetragonal phase occurs and the polar modes appear in the spectra. In a temperature range of 180-230 K, the BST0.5 film undergoes the transition to the monoclinic *r* phase, which agrees well with the thermodynamic theory, developed for BST*x* films in [5].

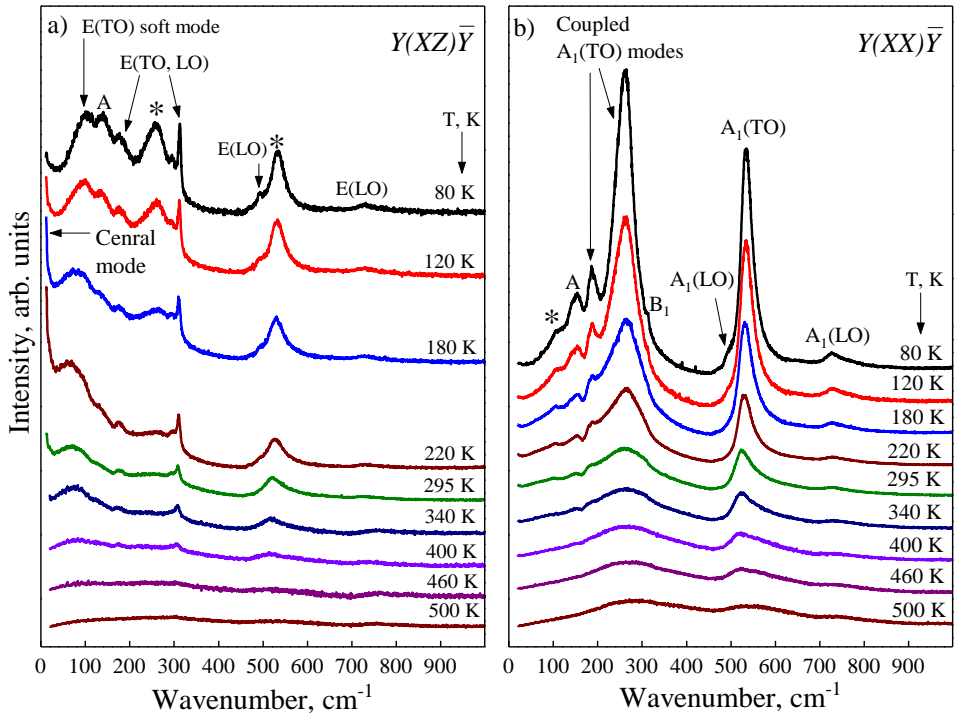

**Figure 3**. Raman spectra of the bi-color BT/BST0.5 superlattice in the (a) crossed and (b) parallel scattering geometries at selected temperatures. Phonon modes in the spectra, leaked from other scattering geometries due to depolarization are marked by the asterisks. The A, B, C and D bands are disorder-induced lines.

Raman spectra of the bi-color BT/BST0.5 and tri-color BT/BST0.5/ST superlattices in a temperature interval from 80 to 700 K are presented in Fig. 3 and 4 for two scattering geometries, $Y(XZ)\bar{Y}$ and $Y(XX)\bar{Y}$. As can be seen, Raman spectra are well polarized in the whole temperature range. Detailed studies of the BT/BST0.5 superlattice in a whole temperature range and BT/BST0.5/ST in the high-temperature range are given in Refs. [9] and [12]. Now we perform the comparative analysis of the low-frequency Raman spectra for the single-component BST0.5 film and bi- and tri-color superlattices with the alternating BT, BST0.5, and ST layers in a broad temperature interval.

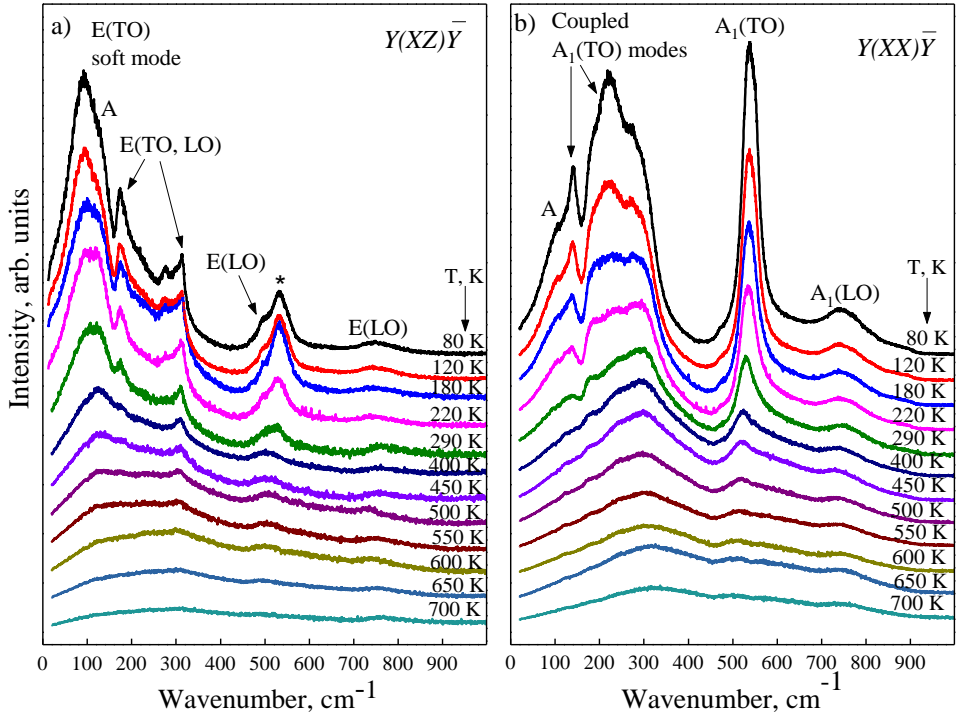

**Figure 4**. Raman spectra of the tri-color BT/BST0.5/ST superlattice in the (a) crossed and (b) parallel scattering geometries at selected temperatures. The $A_1$-type phonon modes in $Y(XZ)\bar{Y}$ spectra, leaked from diagonal geometries due to depolarization are marked by the asterisks. The A, B, C and D bands are the disorder-induced lines.

Raman response of the BT/BST0.5 and BT/BST0.5/ST superlattices in the paraelectric phase is similar to that in BT crystal. Two broad bands at 260 and 530 cm$^{-1}$ have been observed in BT crystal due to disorder in the cubic phase [14]. In the bi- and tri-color superlattices similar bands were observed at 285 and 540 cm$^{-1}$. Such frequency change suggests that the Ba/Sr-substitution modifies the potential relief of the Ti ions in BST$x$ layers, correspondingly to the eight-site model [22]. Below Curie temperature, ~500 K for BT/BST0.5 and ~600 K for BT/BST0.5/ST, the narrow lines of the ferroelectric phase appear at the background. The intensity of the polar modes steadily increases on cooling, while the intensity of the disorder-induced bands, A, B, C, and D, becomes negligibly small with decreasing of the temperature.

It worth to note, that the frequencies of all polar modes in the Raman spectra have no remarkable changes with temperature. Contrariwise, the $Y(XZ)\bar{Y}$ spectra exhibits changes in the low-frequency interval of about 50-100 cm$^{-1}$. The intense low-frequency soft mode is clearly seen at around 50-100 cm$^{-1}$ in the $Y(XZ)\bar{Y}$ spectra of the BT/BST0.5 and BT/BST0.5/ST superlattices, corresponding to E modes in the case of the tetragonal symmetry. The behavior of the E(TO) soft mode observed in the $Y(XZ)\bar{Y}$ spectra depends on the epitaxial stresses of the heterestructures. At room temperature, the frequencies of the E(TO) soft modes of the bi-color BT/BST0.5 and tri-color BT/BST0.5/ST superlattice are about of 102 and 92 cm$^{-1}$, respectively. We suppose that the significant shift of the soft mode frequency demonstrates the increase in the two-dimensional stresses in BT/BST0.5/ST because of the two-dimensional clamping of the BT layers by the ST layers caused by the lattice mismatch between BT and ST layers. According to the XRD data, the unit cell parameter $c$ of the BT and BST layers in the bi-color BT/BST0.5 superlattice is 0.4022 and 0.3948 nm, respectively. The modulation period $\Lambda$ of the bi-color BT/BST0.5 superlattice was found to be about 130 Å [9]. In the tri-color BT/BST0.5/ST superlattice, the out-of-plane $c$ parameters of the BT, BST, and ST layers differ significantly and consist $c_{BT} = 0.4080$ nm, $c_{BST} = 0.3935$ nm, and $c_{ST} = 0.3926$ nm respectively. The modulation period $\Lambda$ is equal to 150 Å [9]. Thereby, the tri-color BT/BST0.5/ST superlattice exhibits a large increase in the degree of tetragonality of the BT layer compared to that, found in the bi-color superlattice. The observed increase of the out-of-plane lattice parameter is associated with the compression of BT layers by the conjugate ST layers. This effect leads to the decreasing of the frequency of the soft mode E(TO).

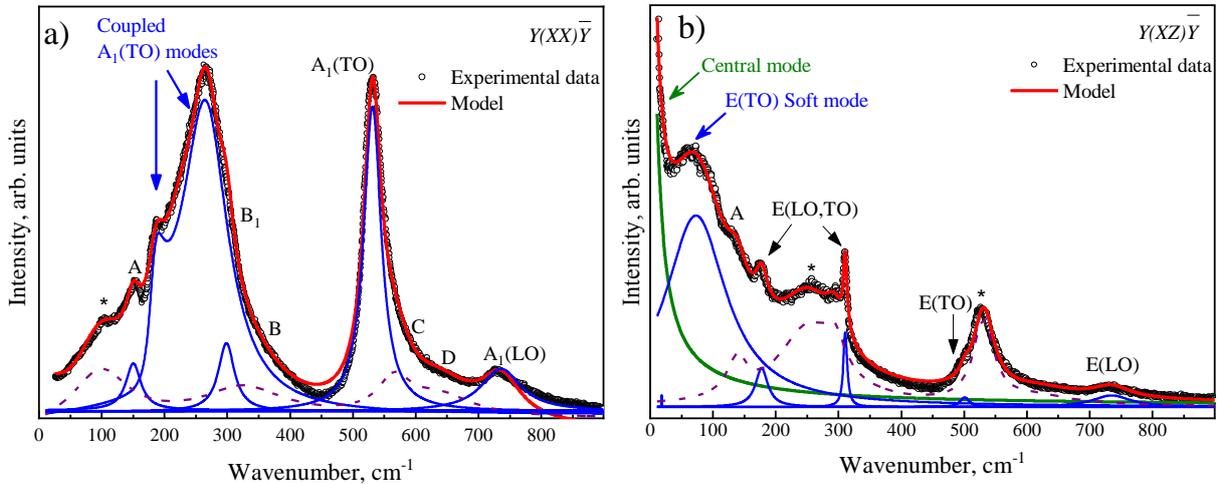

**Figure 5**. Raman spectra of the BT/BST0.5 superlattice in the (a) parallel and (b) crossed scattering geometries at 200 K and the results of their deconvolution. The A$_1$ and B$_1$ modes observed in the $Y(XX)\bar{Y}$ spectra and the E-type eigenmodes in the $Y(XZ)\bar{Y}$ spectra are denoted by blue solid lines. The Debye relaxator is shown by green solid line. The leaked A$_1$ phonon modes and additional disorder-activated A, B, C and D modes are marked by dashed lines.

As known for BT crystal [14], the A$_1$ modes in the $Y(XX)\bar{Y}$ spectra cannot be approximated using a model of additive harmonic oscillators. Consequently, in this case, we used a model of coupled harmonic oscillators [25]. The result of approximation of the Raman spectra of the BT/BST0.5 superlattice in the parallel and crossed scattering geometries is presented in Fig. 5a, b. as separate contours for each phonon mode. The Figure 6a shows the result of deconvolution of the $Y(XZ)\bar{Y}$ spectrum for tri-color superlattice.

In addition to the soft mode band, a pronounced Debye-type central mode is clearly seen in $Y(XZ)\bar{Y}$ the spectrum of the bi-color BT/BST0.5 superlattice, as shown in Fig. 5b. In agreement with earlier studies [26], this central peak is attributed to the relaxational motion of the off-center Ti ion in the plane, parallel to the substrate surface. It is also worth to note that we have not detected a central peak in the $Y(XX)\bar{Y}$ spectra with $A_1$-symmetry modes (Fig. 5a) attributed to a dynamic disorder of the Ti ions along the polar $c$ axis.

Only the coupled $A_1$-type eigenmodes at frequencies of 186 and 264 cm$^{-1}$, $B_1$ mode at about 299 cm$^{-1}$, and disorder-induced bands are observed in the parallel scattering geometry (Figs. 3b and 5a). Moreover, we do not observe the central peak at sub-THz frequency range in the Raman spectra of the BST0.5 film and tri-color BT/BST0.5/ST superlattice (Figs. 2a and 6). The E(TO) soft mode parameters in the spectra of tri-color BT/BST0.5/ST superlattice as a function of temperature are given in Fig. 6b. Fig. 6b show that the soft mode frequency increases from 125 to 138 cm$^{-1}$ with simultaneous increasing of half-width from 125 to 177 cm$^{-1}$ when temperature increases from 80 up to 700 K. The macroscopic phase transition in tri-color BT/BST0.5/ST superlattice occurs at a temperature of around 600 K.

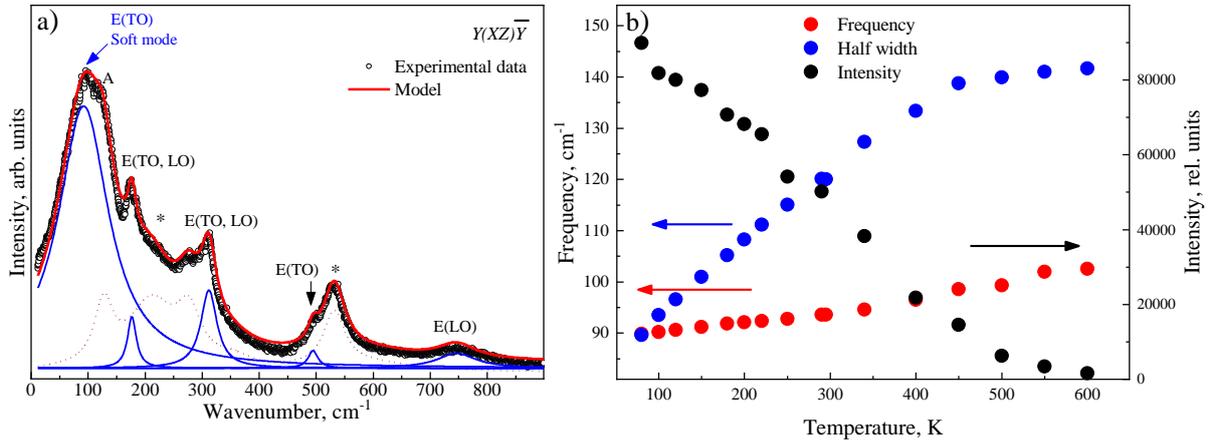

**Figure 6**. (a) Raman spectrum of the BT/BST0.5/ST superlattice in the $Y(XZ)\bar{Y}$ geometry at 200 K and the result of its deconvolution. The E-modes are denoted by a blue solid line. The leaked $A_1$ phonon modes and additional disorder-activated A, B, C and D modes are marked by dashed lines; (b) The temperature dependences of integral intensity, frequency and the half-width of the soft mode in the tri-color BT/BST0.5/ST superlattice.

The temperature evolution of the parameters of the soft and central modes of the bi-color superlattice obtained from the fit $Y(XZ)\overline{Y}$ of the spectra is presented in Fig. 7. The coexistence of the soft and central modes in the $Y(XZ)\overline{Y}$ spectra of the bi-color superlattice was revealed in the temperature range of 120-400 K (Fig. 3). The soft mode appears in the spectra of the bi-color superlattice at ~500 K, suggesting that the transition to ferroelectric phase is upshifted as compared with the BT crystal ($T_C$ ~ 398 K). On further cooling, the intensity of the soft mode increases and an additional central mode appears below ~400 K [12]. The temperature dependencies of the intensity (strength) of the soft and central modes show two broad maxima at around 280 K and 200 K (Fig. 7 a,b).

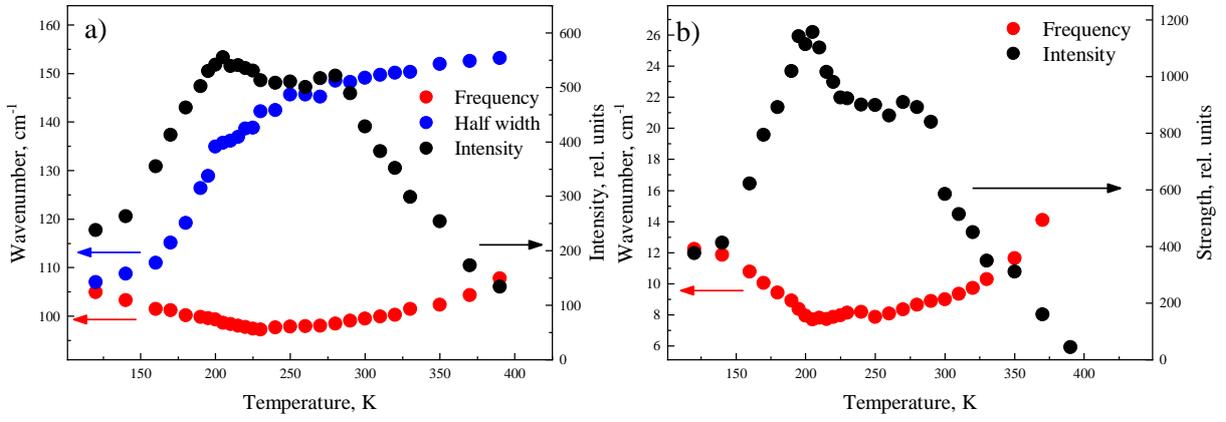

**Figure 7**. Temperature dependences of (a) integral intensity, frequency and half-width of the soft mode, and (b) strength and frequency of the central mode in the spectra of the bi-color BT/BST0.5 superlattice.

As shown in Fig. 7 a,b, the frequencies of the soft mode and Debye relaxator modes decrease on cooling, approaching to rather flat minima at around 200-300 K. Below 200 K they exhibit hardening. The intensity of the soft mode and the strength of the central peak demonstrate two broad maxima at ~280 and ~200 K. This behavior of the soft and central modes can be interpreted as the series of the phase transitions occurred in the alternating BT and BST0.5 layers of the bi-color BT/BST0.5 superlattice [12]. Thus, the BT layers undergo $p - c$ transition at 430 K, and $c - r$ transition at around 270-280 K. Otherwise, in the BST0.5 layers the interval 270-280 K corresponds to the $p - aa$ transition, and the interval 200-210 K to the $c - r$ transition.

## 4. Conclusions

We performed the comparative Raman studies of the BST0.5 film and the BT/BST0.5 and BT/BST0.5/ST superlattices with close modulation periods. It was defined that the BST0.5 film undergoes ferroelectric phase transitions below room temperature. An analysis of the Raman spectra shows that, at room temperature, the soft mode *E*(TO) frequency is significantly higher in the bi-color BT/BST0.5 superlattice than that in the tri-color BT/BST0.5/ST superlattice. In the tri-color superlattice, the two-dimensional deformation of the epitaxial layers extends the stability region of the ferroelectric phase up to ~600 K. Based on the temperature dependence of the Raman spectra, it was established that the two-layer BT/BST0.5 superlattice undergoes the phase transition to the paraelectric phase at the temperature of about 500 K. Moreover, it was observed that the spectra of the BT/BST0.5 superlattice contain a prominent central mode in addition to soft mode at sub-THz frequency range. In bi-color superlattice the soft mode and Debye-type relaxator coexist in a broad temperature range 120-400 K, that indicates the complicated combination of displacement and order-disorder type phase transitions. Such type of behavior leads to the relaxor-like behavior of the dielectric permittivity. The results were interpreted in the frame of the eight-site model [21] for the single-domain state although the sub-Terahertz vibrations of domain walls [27], induced by the depolarization effects in superlattices [28] can also contribute to the spectra. We demonstrated that fabrication of the BT-based heterostructures with appropriate engineering of the internal strain allows for tuning the parameters of their low-frequency phonon spectra in the vast parameter range that can be suitable for the foreseen opto- and radioelectronic applications.

**Acknowledgment**

This work was supported by the Russian Foundation for Basic Research, project No. 17-02-01247a, and by H2020-RISE-ENGIMA action (IAL and AGR). AGR thanks grant of President RF No. SP-1359.2016.3.